\documentclass[a4paper,11pt]{article}
\pdfoutput=1 % if your are submitting a pdflatex (i.e. if you have
\usepackage{jheppub} % for details on the use of the package, please
\usepackage[T1]{fontenc} % if needed
\usepackage{braket}
\usepackage[dvipsnames]{xcolor}
\usepackage{graphicx}
\usepackage{amsmath}

\title{\boldmath Generalized Euler Equation from Effective Action: Implications for the Smarr Formula in AdS Black Holes}

\author{Robinson Mancilla}
\affiliation{
Department of Physics, University of California, Santa Barbara, CA 93101, United States}
\emailAdd{rhmancilla@ucsb.edu}
\abstract{We derive a generalized Euler equation, $\epsilon+p=sT+\mu q+y\frac{\partial p}{\partial y}$, using the effective field theory formulation of perfect fluids. This generalization was achieved by introducing a new variable $y$ into the effective action, which encodes a geometrical scale of the spacetime where the fluid is on. Notably, the generalized Euler equation is independent of the AdS/CFT correspondence. However, when applied to a holographic perfect fluid, this equation naturally recovers the Smarr formula for AdS black holes, thus situating the physical interpretation of the Smarr formula within the framework of well-established physics. Finally, our findings raise important questions regarding the validity of treating the cosmological constant $\Lambda$ as a thermodynamic variable, as proposed in certain frameworks within the literature.}

\begin{document} 
\maketitle
\flushbottom

\section{Introduction}

%%%%%%%%%% Section organization %%%%%%

The AdS/CFT correspondence \cite{Juan1} has greatly advanced our understanding of quantum field theory in the strong coupling regime, in part by geometrizing abstract concepts from quantum physics in gravitational terms. For instance, a thermal state in field theory is dual to an eternal AdS black hole \cite{Juan2}. Moreover, the Hawking-Page transition between thermal AdS spacetime and a "large" AdS black hole \cite{Hawking} is interpreted as the confinement/deconfinement transition in a gauge theory on $S^1 \times S^3$ \cite{Witten}. Within this holographic framework, \cite{Karch} derived the Smarr formula for AdS black holes, providing further insight into its thermodynamic interpretation.

In this article, we revisit the physical meaning of the Smarr formula within the context of gauge/gravity duality. Our goal is to explore the Smarr formula from the perspective of fluid/gravity duality \cite{Hubeny1, Ranga, Hubeny2}. It is well-established that the boundary stress tensor of many AdS black holes takes the form of a perfect fluid in global equilibrium \cite{Clifford0, Clifford}. From this perspective, the thermodynamics of the holographic fluid directly corresponds to black hole thermodynamics. This connection establishes a natural bridge between the Smarr formula and the theory of perfect fluids, which we explore in detail.

A significant development relevant to this work is the recent construction of effective actions for perfect fluids \cite{Dubovsky1, Son, Dubovsky2, Liu2, Liu3, Ranga1, Ranga2, Ranga3, Ranga4, Jensen1}. Within this framework, the effective action not only reproduces the perfect fluid stress tensor but also ensures the validity of the Euler equation, with the effective Lagrangian acting as a local thermodynamic potential. As we will explore in Section \ref{Sec:Smarr formula}, the Smarr formula, when suitably reformulated, bears a close resemblance to the Euler equation — with an additional correction term. This resemblance motivates the pursuit of a generalized Euler equation, which we establish via an explicit derivation from the effective action for perfect fluids. Crucially, this derivation introduces a new variable into the local thermodynamic potential — one that is invariant under both spacetime diffeomorphisms and the emergent low-energy gauge symmetries of the fluid. As we will show, this variable admits a natural geometric interpretation and leads directly to the central result of this work:
\begin{equation}\label{localSmarrLaw0}
    \epsilon+p=sT+\mu q+y\frac{\partial p}{\partial y}~.
\end{equation}
We refer to this as the generalized Euler equation, with the additional term $y\frac{\partial p}{\partial y}$ representing an extra contribution mentioned earlier. Importantly, this result is independent of black hole physics and the AdS/CFT correspondence. We then apply this equation to the case of a holographic perfect fluid, from which the Smarr formula is naturally reproduced.

Another prominent context in which the Smarr formula has been extensively discussed is extended black hole thermodynamics (EBHT). This framework provides a specific interpretation of the Smarr formula by introducing a thermodynamic variable associated with the cosmological constant $\Lambda$ \cite{Kastor, Gibbons, Dolan1, Dolan2, Dolan3}. Initially, this variable was identified as a pressure term, $p_{\Lambda} \sim \frac{1}{l^2}$. However, more recent formulations of EBHT reinterpret it as the central charge, $C \sim N^2$, where $N$ represents the number of degrees of freedom in the dual field theory \cite{Visser1, Gao, Visser2, Mann6}. This shift was primarily motivated by \cite{Karch}, which aimed to align EBHT with holography. However, the central result of this article, (\ref{localSmarrLaw0}), challenges these interpretations of the Smarr formula, as we shall discuss later.

This paper is organized as follows: In Section \ref{Sec:Smarr formula}, we review the Smarr formula and highlight its preliminary similarities with the Euler equation. In Section \ref{Sec:EFT Action}, we examine the standard construction of the effective action for perfect fluids and introduce a new variable into the effective action, which enables us to derive a generalized Euler equation. In Section \ref{Illustration}, we demonstrate that the Smarr formula is a specific instance of the generalized Euler equation. Section \ref{sec:EBHT} discusses why our results are inconsistent with the proposal for extended black hole thermodynamics. Finally, in Section \ref{Discussion}, we present our conclusions and suggest potential directions for future research.

%%%%%%%%%%%%%%%%%%%%%%%%%%%%%%%%%%%%%%%%%%
\section{Smarr formula for AdS black holes}\label{Sec:Smarr formula}
%%%%%%%%%%%%%%%%%%%%%%%%%%%%%%%%%%%%%%%%%%
In this section, we will review the Smarr formula for the AdS$_d$ Reissner-Nordström black hole, closely following the conventions in \cite{Karch}. The Einstein-Maxwell action, with a negative cosmological constant, is given by 
\begin{equation}
    S=\frac{1}{16\pi G_N}\int d^dx \sqrt{-g}\left(R-2\Lambda -F^2\right)~,
\end{equation}
where $G_N$ is the $d$-dimensional Newton's constant. The negative cosmological constant $\Lambda$ can be expressed in terms of the AdS radius $l$ as follows: $\Lambda=-\frac{(d-2)(d-1)}{2l^2}$. For the spherical AdS$_d$ Reissner-Nordström black hole with $d>3$, the metric and gauge field are, respectively, 
\begin{equation}
    ds^2=-f(r)dt^2+f(r)^{-1}dr^2+r^2d\Omega_{d-2}^2~, \quad f(r)=\frac{r^2}{l^2}+1-\frac{\tilde{m}}{r^{d-3}}+\frac{\tilde{q}^2}{r^{2d-6}}~,
\end{equation}
\begin{equation}\label{GaugeFieldA}
   A=\left(\mu l-\frac{1}{\gamma}\frac{\tilde{q}}{r^{d-3}}\right)dt~, \quad     \gamma=\sqrt{\frac{2(d-3)}{(d-2)}}~.
\end{equation}
The constant $\mu$ is the chemical potential, and it is defined such that the gauge field $A$ vanishes at the event horizon located at $r_h$
\begin{equation}\label{ChemPot}
    \mu=\frac{1}{\gamma}\frac{\tilde{q}}{l r^{d-3}_h}~.
\end{equation}
The black hole energy $M$ and charge $Q$ are related to the parameters $\tilde{m}$ and $\tilde{q}$ as follows:
\begin{equation}
    M=\frac{(d-2)A_{d-2}\tilde{m}}{16\pi G_N}~, \quad Q=\frac{l \gamma(d-2) A_{d-2}\tilde{q}}{8\pi G_N}~,
\end{equation}
where $A_{D}$ is the area of the $D$-dimensional unit sphere $S^D$. The black hole entropy $S$ and the Hawking temperature $T$ are given, respectively, by
\begin{equation}
    S=\frac{A_{d-2}r_{h}^{d-2}}{4G_N}~,
\end{equation}
\begin{equation}
    T=\frac{(d-1)r_h}{4\pi l^2}+\frac{(d-3)(1-\gamma^2l^2\mu^2)}{4\pi r_h}~.
\end{equation}
The Smarr formula is traditionally derived either by applying Euler's theorem for homogeneous functions to the "bulk" scaling of thermodynamic quantities\footnote{Scaling arguments have been widely used in black hole thermodynamics to derive Smarr formulas, accommodating various matter contents and asymptotic behaviors (see, e.g., \cite{Lu, Alfaro1}).} or by evaluating the Komar integral at spatial infinity and the horizon \cite{Kastor2,Kastor3,Tomas1,Tomas2,Tomas3}. In this work, we adopt a different perspective, interpreting black hole thermodynamics as the thermodynamics of the dual field theory in the strongly interacting regime, in accordance with the AdS/CFT correspondence. Consequently, we derive the Smarr formula using thermodynamic considerations and the Euclidean action \cite{Karch}. To this end, let us consider the grand canonical potential, $\Omega$:
\begin{equation}\label{DOmega}
    \Omega(T,\mu, V)=U-TS-\mu Q~.
\end{equation}
In our context, $M$ represents the internal energy $U$ of the system. For what is coming it is important to note that, in principle, $\Omega$ is a function of the volume $V$, which, in the case of black holes in AdS spacetimes, corresponds to the spatial volume of the boundary metric where the dual field theory resides. For the AdS$_d$ Reissner-Nordström black hole, we compute the grand canonical potential $\Omega$ using the Euclidean on-shell action
\begin{equation}\label{Omega2}
    \Omega=\frac{A_{d-2}}{8\pi G_N}r^{d-3}_{h}-\frac{M}{(d-2)}~.
\end{equation}
Equating the definition of $\Omega$ in (\ref{DOmega}) with the "on-shell value" of $\Omega$ (\ref{Omega2}), we obtain the Smarr formula for AdS black holes: 
\begin{equation}\label{SmarrFormula}
    (d-1)M=(d-2)(TS+\mu Q)+\frac{(d-2)A_{d-2}}{8\pi G_N}r^{d-3}_{h}~.
\end{equation}
In the last term of this expression, the radius of the event horizon $r_h$ should be understood as a function of the thermodynamic variables. In summary, the Smarr formula is an unusual relation among thermodynamic quantities.

Our objective is to develop a deeper physical understanding of the Smarr formula within the broader framework of the AdS/CFT correspondence. In particular, we are going to recast the Smarr formula~(\ref{SmarrFormula}) into a form that naturally generalizes the standard Euler relation, $U + pV = TS + \mu Q$, by incorporating an additional correction term arising from a geometric scale inherent to the system. As previously noted in~\cite{Karch}, this correction captures the fact that the dual field theory lives on a sphere rather than a flat plane. Our analysis will go beyond purely thermodynamic reasoning by incorporating insights from the fluid/gravity correspondence.

In thermodynamics, the Euler equation arises in the thermodynamic limit, where the volume $V$ and the number of particles $N$ both tend to infinity while maintaining a fixed ratio $\frac{N}{V}$. In this limit, the grand canonical potential becomes $\Omega=-pV$, and the Euler equation follows as  $-pV=U-TS-\mu Q$. We seek to extend this reasoning to the Smarr formula. However, in the context of black hole thermodynamics, defining the thermodynamic volume $V$ and pressure $p$ is far from straightforward. This is where fluid/gravity duality offers crucial insight. To define the pressure $p$ for AdS black holes, we observe that the boundary stress-energy tensor associated with these black holes takes the form of a perfect fluid stress tensor \cite{Clifford0, Clifford}. The boundary stress tensor for any given gravitational solution can be calculated using the formula provided in \cite{Kraus,Clifford1,Skenderis}: 
\begin{equation}\label{boundaryStress}
    T_{ab}=\frac{1}{8\pi G_N}\left(h_{ab}\left(K-\frac{(d-2)}{l}\right)+K_{ab}+\frac{lG_{ab}}{(d-3)}+...\right)~,
\end{equation}
where $h_{ab}$ is the induced metric, $K_{ab}$ is the extrinsic curvature, and $G_{ab}$ is the Einstein tensor for a fixed value of the radial direction $r$. The counterterms explicitly written in (\ref{boundaryStress}), are valid for $d<6$; for higher dimensions, see \cite{Clifford1}. As a reminder, $d$ is the bulk dimension, so $(d-1)$ is the boundary dimension. In our case, the boundary metric is
\begin{equation}
    ds^2|_{\partial \mathcal{M}}= -dt^2+l^2d\Omega_{d-2}^2~.
\end{equation}
Thus, the volume of the spatial section of the boundary spacetime is $V=l^{d-2}A_{d-2}$ where the AdS radius $l$ is acting as the radius of the $(d-2)$-dimensional sphere. For even-dimensional boundary spacetime, the conformal anomaly and Casimir energy appear \cite{Kraus,Skenderis}. Since these effects are not relevant to the present discussion, we restrict our analysis to odd-dimensional boundary spacetime. For $(d-1) = 2n + 1$ dimensions, the components of the boundary stress tensor are
\begin{equation}
    -T^t_{\;t}=\frac{(d-2)\tilde{m}}{16\pi G_n l^{d-2}}=\frac{M}{l^{d-2}A_{d-2}}=\frac{M}{V}~,
\end{equation}
\begin{equation}
    T^{\phi_j}_{\;\phi_j}=\frac{\tilde{m}}{16\pi G_n l^{d-2}}=\frac{M}{(d-2)V}~.
\end{equation}
where we denoted by $\phi_j$ an arbitrary angle of the sphere. We can recast these components of $T_{ab}$ as a perfect fluid stress tensor
\begin{equation}
    T_{ab}=p g_{ab}+(p+\epsilon)u_{a}u_{b}~,
\end{equation}
where $g_{ab}$ denotes the boundary metric, and $u_a$ is a unit timelike vector. The pressure $p$ and the energy density $\epsilon$ are
\begin{equation}
    p=\frac{\epsilon}{(d-2)}~, \quad \epsilon=\frac{M}{V}~.
\end{equation}
This holographic perfect fluid in certain sense is dual to the AdS$_d$ Reissner-Nordström black hole. Clearly, this fluid satisfies the conformal equation of state $\epsilon=(d-2)p$. The thermodynamics of this fluid is the same than the black hole thermodynamics since the fluid is at global thermal equilibrium.

Returning to the Smarr formula, substituting the conformal equation of state $M = (d-2)pV$ into (\ref{SmarrFormula}) provides a more suggesting expression:
\begin{equation}\label{SmarrGlobal1}
    M+pV=TS+\mu Q+\frac{A_{d-2}}{8\pi G_N}r^{d-3}_{h}~,
\end{equation}
which resembles the Euler equation plus an additional correction term. As mentioned earlier, the traditional Euler equation arises in the thermodynamic limit, where $V \to \infty$. However, in our case, the volume $V$ occupied by the fluid is finite, and thus the standard derivation of the Euler equation no longer applies. In the gravitational context, the limit $V \to \infty$ corresponds to a charged black brane, for which the Smarr formula reduces to the Euler equation $p+\epsilon=Ts+\mu q$ \cite{Makoto}, as the dual field theory lives in a plane. 

In the analysis leading to equation (\ref{SmarrGlobal1}), we relied on the idea that a perfect fluid in global thermal equilibrium provides an equivalent description of global thermodynamics. However, perfect fluids are typically used to describe local thermodynamics, where the Euler equation is assumed to hold universally—a point that seems at odds with our intuitive understanding of the Smarr formula developed in this section. To fully grasp the physical significance of the Smarr formula, it is essential to extend the Euler equation beyond its conventional domain. Therefore, in the next section, we will set aside the discussion of black holes and focus solely on the Euler equation and its generalization in the context of perfect fluids.

%%%%%%%%%%%%%%%%%%%%%%%%%%%%%%%%%%%%%%%
\section{Effective action for perfect fluids}\label{Sec:EFT Action}
%%%%%%%%%%%%%%%%%%%%%%%%%%%%%%%%%%%%%%%

In this section, we begin by reviewing the standard construction of the effective action for perfect fluids, which naturally leads to the Euler equation. We then extend this framework by introducing a new variable into the effective action, capturing the fluid's response to the presence of a geometrical scale. With this extension, we derive a generalized form of the Euler equation.

\subsection{Review standard effective action for perfect fluids}\label{sec:3.1}

Formal constructions of effective actions for hydrodynamics have been developed using techniques such as the Schwinger-Keldysh formalism \cite{Liu2, Liu3, Ranga1, Ranga2} and superspace methods \cite{Ranga3, Ranga4, Jensen1}. Comprehensive reviews can be found in \cite{Liu4, Basar}. Although these advanced techniques will not be explicitly employed in this article, they highlight the significant progress and depth this field has achieved. Instead, we revisit the foundational constructions presented in \cite{Dubovsky1, Son, Dubovsky2}, informed by a more modern understanding of the subject. Moreover, we focus exclusively on perfect fluids, leaving aside considerations of dissipation and hydrodynamic anomalies. 

Hydrodynamics provides a valid effective description in the long-wavelength limit. For this reason, we assume that the curvature of the manifolds under consideration is much larger than the mean free path of the fluid. To construct the effective action for perfect fluids, two key components are required:

\begin{itemize}
    \item \textbf{Auxiliary Scalar Fields:} To ensure that the covariant conservation of the stress tensor becomes a second-order equation (as it is initially a first-order equation with respect to the fluid variables), auxiliary scalar fields are introduced as the low-energy dynamical variables of the effective action.
    \item \textbf{Symmetry Organization:} To organize the potential terms that can appear in the action principle, the symmetries of the fluid must be known. In this case, the relevant symmetries are low-energy gauge symmetries, which imply that the auxiliary scalar fields can be interpreted as Stueckelberg fields \cite{Liu2, Liu4}.
\end{itemize}
Given the dynamical variables and symmetries of the effective action, one should, in principle, include all gauge-invariant terms organized in a derivative expansion. Since our focus is restricted to perfect fluids, however, we will neglect the higher-derivative contributions that encode dissipative effects. Moreover, as stated in the introduction, we assume that the bulk spacetime has d dimensions. Consequently, in this section, we consider a perfect fluid in an odd $(d-1)=2n+1$ dimensional manifold, ensuring that anomalies and Casimir energy do not play a role in our discussion. Finally, we limit this section to field-theoretical aspects, deferring the connection between these results and gravity to the next section.

Let’s begin by establishing the notation: the background metric for the perfect fluid is
\begin{equation}
    ds^2=g_{ab}dx^adx^b~, \quad a,b \in \{0,1,...,d-2\}~.
\end{equation}
The low-energy dynamical variables for the effective action are $(d-2)$ spatial scalar fields $\sigma^I$ that describe the fluid element, plus a temporal scalar field $\sigma^0$ that acts as an internal clock for the fluid element:
\begin{equation}
    \sigma^I=\sigma^I(x^a)~, \quad \sigma^0=\sigma^0(x^a)~, \quad I \in \{1,...,d-2\}~.
\end{equation}
The fluid dynamics possess the following emergent low-energy internal symmetries: 
\begin{equation}\label{TemporalShift}
 \sigma^0\to\sigma^0+f^0(\sigma^I)~,
\end{equation}
\begin{equation}\label{spatialDiffeo}
    \sigma^I\to f^I(\sigma^J)~.
\end{equation}
These symmetry transformations can be motivated by the observation that $\sigma^I$ represents spatial coordinates describing the fluids. As such, $\sigma^I$ must satisfy the comoving condition\footnote{In the literature, this condition defined the Lagrangian observer.} 
\begin{equation}\label{Comoving Condition}
    \frac{d }{d\tau}\sigma^I(x^a(\tau))=0 \Longleftrightarrow u^a\partial_a \sigma^I=0~,
\end{equation}
where $x^a$ in the position in the spacetime, and the fluid velocity is $u^a=\frac{dx^a}{d\tau}$. The second expression in (\ref{Comoving Condition}) employs the directional derivative along the fluid velocity $u^a$. Notably, this condition is preserved by the symmetry transformations (\ref{TemporalShift}) and (\ref{spatialDiffeo}). It is also important to highlight that, unlike $\sigma^I$, the temporal scalar field satisfies $\frac{d}{d\tau} \sigma^0(x^a(\tau)) \neq 0$, yet it remains gauge-invariant under these symmetry transformations.  

In essence, the perfect fluid is characterized by a target metric $G_{MN}(\sigma^M)$, where the fluid coordinates are $\sigma^M = (\sigma^0, \sigma^I)$. In this framework, the internal symmetries (\ref{TemporalShift}) and (\ref{spatialDiffeo}) correspond to the isometries of a degenerate metric \cite{Son}:
\begin{equation}\label{TargetMetric}
    ds^2=G_{MN}d\sigma^Md\sigma^N~, \quad G_{NM}n^M=0~, \quad M,N \in \{0,1,...,d-2\}~,
\end{equation}
where $n^N$ is a null-like eigenvector associated with the zero eigenvalue of $G_{MN}$. An explicit parameterization of this degenerate target metric in terms of spatial target metric\footnote{In the context of fluid/gravity duality, the target metric $G_{MN}$ is essentially the horizon metric. In this way, the horizon asymptotic symmetry becomes the emergent low-energy gauge symmetries of the effective action for hydrodynamics \cite{Liu5}.} is 
\begin{equation}\label{SpatialG}
    ds^2=0 \cdot (d\sigma^0)^2+G_{IJ}(d\sigma^I-v^I d\sigma^0)(d\sigma^J-v^J d\sigma^0)~, \quad I,J \in \{1,...,d-2\}~,
\end{equation}
where $v^J$ denotes boosts. The null normal vector $n^M$ is then given by
\begin{equation}
    n^M=\frac{1}{\gamma}(1,v^J)~, \quad \gamma^{-1}=\sqrt{1-G_{IJ}v^Iv^J}~.
\end{equation}
Thus, the fluid coordinates $\sigma^M$ are Stueckelberg fields associated with a sub-class of diffeomorphisms of the target metric. Further discussion of this point is found in \cite{Liu2, Liu4}. 

So far, we have provided the description for neutral fluids. To describe charged fluids, a Stueckelberg field associated with a $U(1)$ symmetry must be introduced \cite{Dubovsky2}
\begin{equation}
    \varphi=\varphi(x^a)~,
\end{equation}
with the following internal transformation, dubbed chemical shift symmetry:
\begin{equation}\label{GaugeShift}
    \varphi\to\varphi+g(\sigma^I)~.
\end{equation}
Given the low-energy dynamical variables $\sigma^M$ and $\varphi$, along with the corresponding low-energy gauge symmetries, we proceed to construct the effective action for a charged perfect fluid. To achieve this, we must identify scalar quantities that remain invariant under diffeomorphisms of the physical metric $g_{ab}$ and are also gauge-invariant under the fluid symmetry transformations (\ref{TemporalShift}) and (\ref{spatialDiffeo}).

One such scalar is the entropy density $s$ of the system, which is defined in terms of the spatial target metric $G_{IJ}$ (\ref{SpatialG}) as follows \cite{Son}:
\begin{equation}\label{entropy}
    s=\tau_0\sqrt{\text{Det}(B^{IJ}G_{JK})}~, \quad B^{IJ}=g^{ab}e_{a}^{I}e_{b}^{J}~,
\end{equation}
\begin{equation}
    e_{a}^{I}\equiv\partial_{a}\sigma^{I}-v^I\partial_{a}\sigma^0~.
\end{equation}
Here, $\tau_0$ is a dimensionful constant that provides the correct unit for the entropy density.
Another scalar is the chemical potential $\mu$ associated with the $U(1)$ symmetries \cite{Dubovsky2} which reads
\begin{equation}\label{ChemicalPot}
    \mu=u^{a}\partial_{a}\varphi~.
\end{equation}
We note that both $s$ and $\mu$ are local functions of the physical spacetime where fluid is on, and they take constant values only at global thermal equilibrium. 

The low-energy fields $\sigma^M$, $\varphi$ are auxiliary variables that allow the description of perfect fluids and their local thermodynamics in terms of an action principle. To zero-order in the derivative expansion, the effective action for a perfect fluid is:
\begin{equation}\label{EFT1}
    I=-\int d^{d-1}x \sqrt{-g}\zeta(s,\mu)~,
\end{equation}
where $\zeta$ is a local thermodynamic potential function of the entropy density $s$ and the chemical potential $\mu$. By taking the variation with respect to the background metric $g_{ab}$, one obtains the stress tensor:
\begin{equation}\label{StressTensor11}
    T_{ab}=\left(-\zeta+s\frac{\partial\zeta}{\partial s}\right)g_{ab}+\left(-\mu \frac{\partial\zeta}{\partial \mu}+s\frac{\partial\zeta}{\partial s}\right)u_{a}u_{b}~.
\end{equation}
Comparing this result with the stress tensor of a perfect fluid $T_{ab} = p g_{ab} + (p+\epsilon)u_{a}u_{b}$, one deduces the pressure $p$ 
\begin{equation}
    p=-\zeta+s\frac{\partial\zeta}{\partial s}~,
\end{equation}
and the local energy density $\epsilon$
\begin{equation}
    \epsilon=\zeta-\mu \frac{\partial \zeta}{\partial \mu}~.
\end{equation}
The local thermodynamic potential $\zeta$ is essentially a Legendre transformation: $\zeta = \epsilon - \mu q$, where the charge density is $q = -\frac{\partial \zeta}{\partial \mu}$. Then, the temperature is simply $T = -\frac{\partial\zeta}{\partial s}$. In this way, the consistency of the effective action (\ref{EFT1}) and the stress tensor (\ref{StressTensor11}) with the standard thermodynamic relation implies the Euler equation:
\begin{equation}\label{EuleEqutionDensity}
    p+\epsilon=sT+\mu q~.
\end{equation}
Thus, the Euler equation is an unavoidable consequence of the effective action (\ref{EFT1}). Note that no equation of state for the fluid is used at all in this derivation.

For completeness, we mention that due to the comoving condition (\ref{Comoving Condition}), the entropy current $s^a$ is conserved off-shell:
\begin{equation}\label{eq:entropy conservation}
    \nabla_a s^a=0 , \quad s^a=s u^a~.
\end{equation}
This conservation law reflects the non-dissipative nature of the fluid. In addition, the chemical shift symmetry (\ref{GaugeShift}) implies the on-shell conservation of the charge current:
\begin{equation}\label{eq:U(1)conservation}
    \nabla_a q^a=0 , \quad  q^a=q u^a~.
\end{equation}
Finally, the entropy current can be rewritten using the Euler equation \eqref{EuleEqutionDensity} as
\begin{equation}\label{eq:currentFormalism}
    s^a=p\beta^a-T^{ab}\beta_b-\frac{\mu}{T} q^a~,
\end{equation}
where $\beta^a\equiv \frac{u^a}{T}$. This expression for $s^a$ can serve as a starting point for the current formalism of fluids \cite{Ranga5}. With this, we conclude our review of the effective action formulation for perfect fluids. In the following section, we will use this framework to derive a generalized Euler equation.

%%%%%%%%%%%%%%%%%%%%%%%%%%%%%%%%%%%%%%%%%%%%%%%%%%%%%%
\subsection{Generalized Euler equation}\label{sec:3.2}
%%%%%%%%%%%%%%%%%%%%%%%%%%%%%%%%%%%%%%%%%%%%%%%%%%%%%%

To generalize the Euler equation, we introduce a new variable into the effective action for perfect fluids. This variable must be a diffeomorphism-invariant scalar with respect to $g_{ab}$ and must remain gauge-invariant under the shift symmetries given in (\ref{TemporalShift}), (\ref{spatialDiffeo}). As discussed at the end of Section \ref{Sec:Smarr formula}, there are scenarios in which the standard scaling argument used to derive the Euler equation fails—for instance, when the volume of the system is compact. In such a case, the radius of the compact space introduces an additional geometric scale into the problem, implying that different thermodynamic quantities respond differently under scale transformations. This observation suggests that the new variable we seek should have a geometric origin and transform appropriately under scaling. The gauge-invariant quantity that satisfies these criteria is
\begin{equation}\label{y}
    y=u^{a}\partial_{a}\sigma^0~.
\end{equation}
The definition of this variable closely resembles the definition of the chemical potential $\mu$ presented in (\ref{ChemicalPot}). However, the temporal scalar field $\sigma^0$, which represents the clock for the fluid element, differs in nature from the Stueckelberg field $\varphi$ associated with U(1) symmetry.\footnote{The variable $y$ has been previously considered in the literature, but \cite{Ballesteros} did not clearly distinguish between the temporal scalar field $\sigma^0$ and the gauge scalar field $\varphi$.} In fact, another way to understand the expression in \eqref{y} is to notice that the pushforward of the fluid velocity is given by the vector
\begin{equation}
    Y^M=u^{a}\partial_{a}\sigma^M
\end{equation}
which lives in the fluid space. However, due to the shift symmetries \eqref{TemporalShift} and \eqref{spatialDiffeo}, only the temporal component of $Y^M$ is non-vanishing. Hence, the expression in \eqref{y} simply reduces to $y\equiv Y^0$. Let us note that the vector $Y^M$  of the fluid space transforms covariantly under arbitrary diffeomorphisms of the target space. Under the restricted shift symmetries, however, its temporal component remains gauge-invariant. Since the target space is a null surface, this makes the variable $y$ a natural quantity to introduce within the fluid EFT. These observations clarify the mathematical origin of $y$; its physical interpretation, on the other hand, is less immediate. We will return to this point at the end of the section, where the physical meaning of $y$ will be analyzed in detail. 

The effective action for perfect fluids is now
\begin{equation}\label{EFTACTION2}
    I=-\int d^{d-1}x \sqrt{-g}\chi(s,\mu,y)~,
\end{equation}
where $\chi$ represents a new local thermodynamic potential that depends on entropy density $s$, chemical potential $\mu$, and the new variable $y$. Taking the variation with respect to the background metric $g_{ab}$, the stress tensor follows
\begin{equation}
    T_{ab}=\left(-\chi+s\frac{\partial\chi}{\partial s}\right)g_{ab}+\left(-y\frac{\partial\chi}{\partial y}-\mu \frac{\partial\chi}{\partial \mu}+s\frac{\partial\chi}{\partial s}\right)u_{a}u_{b}~.
\end{equation}
The next step is to compare this expression with the stress tensor of a perfect fluid: $T_{ab} = p g_{ab} + (p + \epsilon) u_{a} u_{b}$. Thus, we can identify the pressure $p$ and energy density $\epsilon$ as:
\begin{equation}\label{PE}
    p=-\chi+s\frac{\partial\chi}{\partial s}~, \quad \epsilon =\chi-y\frac{\partial \chi}{\partial y}-\mu\frac{\partial \chi}{\partial \mu}~.
\end{equation}
The following combination will be relevant later:
\begin{equation}\label{AlmostSmarr}
     \epsilon +p=-y\frac{\partial \chi}{\partial y}-\mu\frac{\partial \chi}{\partial \mu}+s\frac{\partial\chi}{\partial s}~.
\end{equation}
As a second step, we need to properly identify the rest of the thermodynamic variables. We note that the energy density has the following dependence: $\epsilon = \epsilon(s, q, y)$. Thus, if we perform the Legendre transformation $\tilde{\epsilon} = \epsilon - \mu q$, we find that $\tilde{\epsilon} = \tilde{\epsilon}(s, \mu, y)$. In this way, the temperature $T$ and charge density $q$ are given by:
\begin{equation}
    T=\left.\frac{\partial \tilde{\epsilon}}{\partial s}\right|_{y, \mu}~, \quad q=\left.\frac{\partial \tilde{\epsilon}}{\partial \mu}\right|_{y, s}~.
\end{equation}
From these expressions, we deduce how to compute the local temperature $T$ and charge density $q$ from the thermodynamic potential $\chi$:
\begin{equation}\label{LocalTq}
    T=\frac{\partial \chi}{\partial s}-y \frac{\partial^2 \chi}{\partial y \partial s}~, \quad q=-\frac{\partial \chi}{\partial \mu}~.
\end{equation}
Inserting these results into (\ref{AlmostSmarr}), we obtain the relation
\begin{equation}
    \epsilon+p=sT+\mu q+y\partial_y\left(-\chi +s\frac{\partial \chi}{\partial s}\right)~.
\end{equation}
In parentheses of the last term, we recognize the local pressure $p$ given in (\ref{PE}). Therefore, we obtain a new relation among thermodynamic variables:
\begin{equation}\label{SmarrLaw}
    \epsilon+p=sT+\mu q+y\frac{\partial p}{\partial y}~.
\end{equation}
This expression is the generalized Euler equation that we were seeking. The additional correction involves a derivative of the pressure $p$ with respect to $y$. Therefore, if the thermodynamics of a fluid in a given spacetime does not respond to variations in $y$, the additional correction $\frac{\partial p}{\partial y}$ vanishes, recovering the standard Euler identity.

Another important aspect of the generalized effective action for a perfect fluid (\ref{EFTACTION2}) is the emergence of a new conserved current. In addition to the off-shell conservation of the entropy current (\ref{eq:entropy conservation}) and the on-shell conservation of the charge current (\ref{eq:U(1)conservation}), we now also have:
\begin{equation}\label{eq:NEWconservation}
    \nabla_a J_{(y)}^a = 0, \quad J_{(y)}^a = u^a \partial_y p~.
\end{equation}
This conservation law arises from the fact that the variable $y$, defined in (\ref{y}), is invariant under spatial diffeomorphisms of the target space, $\sigma^I \to  f^I(\sigma^J)$. Because of this, the entropy current \eqref{eq:currentFormalism} is now modified to be
\begin{equation}
     s^a=p\beta^a-T^{ab}\beta_b-\frac{\mu}{T} q^a-\frac{y}{T}J^a_{(y)}~.
\end{equation}
To further explore the physical meaning of the variable $y$, let us consider a scenario of global thermal equilibrium, where the fluid clock can be aligned with physical time; that is, $\sigma^0 = x^0$. In this case, we find that $y = u^0$, an expression that may appear somewhat opaque from the physics point of view at first glance. To build intuition, let us perform a Weyl transformation of the physical metric, $g' = W^2 g$, where $W$ is a conformal factor. Under this transformation, the fluid velocity transforms as $u'^a = \frac{u^a}{W}$. This leads to the following relation:
\begin{equation}
    y \to y' =\frac{u^0}{W}~.
\end{equation}
Thus, the variable $y$ is naturally associated with the inverse of the conformal factor $W$. To fully grasp the physical meaning of the variable $y$, let us recall that thermodynamic variables are not Weyl invariant; the introduction of the variable $y$ provides a convenient bookkeeping device to track how these variables transform under scaling transformations. In fact, for constant $y$ the operator $y\partial_y$ appearing in the generalized Euler equation \eqref{SmarrLaw} plays the role of the dilation operator, since $y\partial_y=x^a\partial_a$ in line with the interpretation discussed above. Certainly, for non-constant $y$, the operator $y\partial_y$ should not be identified anymore with the dilation operator. 

To facilitate comparison between the central result of this work and existing literature, we rewrite the generalized Euler identity (\ref{SmarrLaw}) in the following form:
\begin{equation}\label{eq:free energy}
    \omega = \epsilon - sT - \mu q \Longrightarrow \omega = -p + y \frac{\partial p}{\partial y}~,
\end{equation}
where $\omega$ denotes the density of the grand canonical potential. This expression highlights that the presence of an additional scale in the system, encoded by the variable $y$, challenges the conventional assumption that, in the thermodynamic limit, the grand canonical potential $\omega$ is simply proportional to the pressure $p$.

The relation in (\ref{eq:free energy}) bears a notable resemblance to earlier results in the literature on effective actions for fluids \cite{Minwalla, Jensen0}.\footnote{We thank Kristan Jensen for bringing these references to our attention.} However, a key distinction lies in the construction of our effective action, which is formulated at zeroth order in the derivative expansion. In contrast, the works cited in \cite{Minwalla, Jensen0} incorporate higher-order derivative corrections, which modify the form of the free energy accordingly.

\section{Smarr formula as a particular case of the generalized Euler equation}\label{Illustration}

As discussed around equation (\ref{SmarrGlobal1}), the Smarr formula can be viewed as a generalization of the Euler equation. Traditionally, however, the thermodynamics of black holes is examined using Euclidean methods. Unfortunately, these methods do not readily define the thermodynamic volume $V$ and pressure $p$ of a thermal system within the context of gravity. This limitation motivates our use of fluid dynamics to analyze the Smarr formula. We will now demonstrate the generalized Euler equation (\ref{SmarrLaw}), derived in the previous section, reduces to a local version of the Smarr formula when applied to holographic fluids. Subsequently, the standard form of the Smarr formula emerges by considering the fluid in global thermal equilibrium, which encapsulates the same thermal information as the first law of thermodynamics.

In this section, we extend the discussion from section \ref{Sec:Smarr formula} by examining the AdS$_d$ Reissner-Nordström black hole with arbitrary horizon topology. We adopt Eddington-Finkelstein coordinates, as this gauge simplifies the analysis of Weyl transformations compared to the Fefferman-Graham gauge. Moreover, we restrict the boundary dimension $(d-1)$ to be odd; otherwise, the conformal anomaly and the Casimir energy would need to be accounted for in the effective action. The line element is:
\begin{equation}\label{Bulkg}
    ds^2=-f(r)du^2-2dudr+r^2d\Sigma_{k}^2~, \quad f(r)=\frac{r^2}{l^2}+k-\frac{\tilde{m}}{r^{d-3}}+\frac{\tilde{q}^2}{r^{2d-6}}~.
\end{equation}
Here, the gauge field $A$ is still given by equation (\ref{GaugeFieldA}). The constant $k$ represents the unit-normalized curvature of the transverse space $\Sigma_k$. Thus, the black brane is recovered for $k=0$, the spherical black hole for $k=1$, and the hyperbolic black hole for $k=-1$. In all cases, the black hole energy $M$ and charge $Q$ are related to the parameters $\tilde{m}$ and $\tilde{q}$ as follows:
\begin{equation}\label{4.2}
    M=\frac{(d-2)\Sigma_{k,d-2}\tilde{m}}{16\pi G_N}~, \quad Q=\frac{\gamma l (d-2)\Sigma_{k,d-2}\tilde{q}}{8\pi G_N}~.
\end{equation}
The entropy of the black holes is given by:
\begin{equation}\label{4.3}
    S=\frac{\Sigma_{k,d-2}r_{h}^{d-2}}{4G_N}~,
\end{equation}
where $\Sigma_{k,d-2}$ represents the volume of the transverse space with unit radius. It is important to note that for flat and hyperbolic spaces, $\Sigma_{k,d-2}$ is formally infinite. Additionally, hyperbolic space has negative curvature, introducing an extra scale into the physical problem. 

The Hawking temperature $T$ and the chemical potential $\mu$ are
\begin{equation}\label{4.4}
    T=\frac{(d-1)r_h}{4\pi l^2}+\frac{(d-3)(k-l^2\gamma^2\mu^2)}{4\pi r_h}~, \quad \mu=\frac{\tilde{q}}{\gamma l r_{h}^{d-3}}~. 
\end{equation}
Note that the chemical potential $\mu$ is the same as in equation (\ref{ChemPot}). The Smarr formula is now given by:
\begin{equation}\label{SmarrK}
    M+pV=TS+\mu Q+\frac{k\Sigma_{k,d-2}}{8\pi G_N} r^{d-3}_{h}~.
\end{equation}
For $k=0$ the last term vanishes, thereby recovering the standard Euler equation for a charged black brane. This is because, in flat space, there are no additional scales to consider.

We will now compute the thermodynamic potential $\chi$ (\ref{EFTACTION2}) which is associated with the effective action for the holographic perfect fluid dual to the  AdS$_d$ Reissner-Nordström black hole. Our objective is to place the holographic fluid in a boundary spacetime that exhibits spacetime dependence, allowing us to mimic a state that is near out of equilibrium.\footnote{This step is purely for illustrative purposes and carries no additional meaning, as we are disregarding dissipation, and anomalies.} As a consequence, the thermodynamic variables now will become local functions of the spacetime. To obtain the boundary metric with the desired properties, we perform the following radial diffeomorphism in the bulk line element (\ref{Bulkg}):
\begin{equation}\label{RadialDiff}
    r\to W(x^a)r~,
\end{equation}
where $W(x^a)$ is an arbitrary function independent of the radial variable $r$. The new line element is
\begin{equation}
    ds^2=-f(Wr)du^2-2 W dudr-2r\partial_aWdudx^a+W^2r^2d\Sigma_{k}^2~, 
\end{equation}
\begin{equation}
    f(r)=\frac{W^2r^2}{l^2}+k-\frac{\tilde{m}}{(Wr)^{d-3}}+\frac{\tilde{q}^2}{(Wr)^{2d-6}} ~.
\end{equation}
Therefore, the boundary metric is
\begin{equation}\label{BoundaryMetric}
    ds^2|_{\partial \mathcal{M}}=-\frac{W^2}{l^2}du^2+W^2d\Sigma_k^2~.
\end{equation}
This line element describes a non-homogeneous, time-dependent background in which the fluid resides. The radial diffeomorphism (\ref{RadialDiff}) consequently induces a Weyl transformation on the boundary metric,\footnote{In other words, the Weyl transformation is induced by a PBH transformation of the bulk metric. A detailed discussion of the consequences of this transformation in the thermodynamics of asymptotically locally AdS black holes is given in \cite{Ioannis3}.} leading to the following expression for the fluid velocity:
\begin{equation}
    u_a=-\frac{W}{l}\delta^{0}_a~.
\end{equation}
Let's focus now on the target metric $G_{MN}$ that describes the fluid space. Using the unitary gauge embedding $\sigma^0 = x^0$ and $\sigma^I = x^i$ \cite{Son}, the target metric $G_{MN}$ takes a very simple form:
\begin{equation}
    ds^2_{fluid}=-0\cdot du^2+r^2_hd\Sigma^2_k~,
\end{equation}
that matches the horizon metric.\footnote{More precisely, the horizon metric before the Weyl transformation.} Given the target metric and the fluid velocity, we can use the formulas (\ref{entropy}), (\ref{ChemicalPot}), and (\ref{y}) to compute the entropy density $s$, the chemical potential $\mu$, and the variable $y$, respectively:
\begin{equation}\label{s,u,y}
    s=\frac{1}{4G_N}\frac{r_{h}^{d-2}}{W^{d-2}}~, \quad \mu=\frac{\tilde{q}}{\gamma r_h^{d-3}}\frac{1}{W}~, \quad y=\frac{l}{W}~,
\end{equation}
which are the variables that the thermodynamic potential $\chi$ depends on. We note that the entropy density satisfies its conservation law (\ref{eq:entropy conservation}), as expected, and we set $\tau_0 = \frac{1}{4G_N}$ in equation \eqref{entropy}. We will utilize these variables to derive the explicit expression for the new thermodynamic potential $\chi$ from the energy density $\epsilon$ using equation (\ref{PE}). Our next step involves computing the boundary stress tensor via the formula (\ref{boundaryStress}) and the boundary metric given in (\ref{BoundaryMetric}). The components of the stress tensor are 
\begin{equation}
    -T^t_{\;t}=\frac{lM}{W^{d-1}\Sigma_{k,d-2}}~,
\end{equation}
\begin{equation}
    T^{\phi_j}_{\;\phi_j}=\frac{lM}{(d-2)W^{d-1}\Sigma_{k,d-2}}~.
\end{equation}
Similar to section \ref{Sec:Smarr formula}, these components can be expressed in the form of a perfect fluid stress tensor $T_{ab} = p g_{ab} + (p + \epsilon) u_{a} u_{b}$, from which we read the energy density $\epsilon$:
\begin{equation}
     \epsilon=\frac{(d-2)}{16\pi G_N l}\left[\left(4G_N s\right)^{\frac{d-3}{d-2}}(k y^2+(\gamma l)^2\mu^2)+\left(4G_N s\right)^{\frac{d-1}{d-2}}\right]~,
\end{equation}
where $M$ has been expressed in terms of $r_h$ as an intermediate step, and then in terms of the variables $s$, $\mu$, and $y$ given in equation (\ref{s,u,y}). One can also verify that the conformal equation of state: $p = \frac{\epsilon}{d-2}$ is satisfied.

As a final step, using equation (\ref{PE}) that connects $\epsilon$ and $p$ with the thermodynamic potential $\chi$, we can infer the effective action for this holographic fluid
\begin{equation}
    I=-\int d^{d-1}x \sqrt{-g}\chi(s,\mu,y)~,
\end{equation}
with the thermodynamic potential $\chi$ given by
\begin{equation}\label{Chi}
     \chi=\frac{(d-2)}{16\pi G_N l}\left[-\left(4G_N s\right)^{\frac{d-3}{d-2}}(k y^2+(\gamma l)^2\mu^2)+\left(4G_N s\right)^{\frac{d-1}{d-2}}\right]~.
\end{equation}
We note that all thermodynamic variables involved in $\chi$ are local functions of the spacetime where the fluid resides. Certainly, we have derived $\chi$ in this case through inverse engineering, as we already knew some of the thermodynamic variables and the geometrical variable $y$. This approach is not an inconvenience for us; rather, it allows us to treat the thermodynamic potential $\chi$ as a starting point, as it encodes all the thermal information of the fluid. Finally, using equation (\ref{LocalTq}), we compute respectively the local temperature $T$ and charge density $q$ as follows:
\begin{equation}
     T=\frac{(d-1)}{4\pi l}\left(4G_N s\right)^{\frac{1}{d-2}}+\frac{(d-3)(ky^2-(\gamma l)^2\mu^2)}{4\pi l}\left(4G_N s\right)^{\frac{-1}{d-2}}~,
\end{equation}
\begin{equation}
     q= \frac{(d-2)\gamma^2l\mu}{8\pi G_N}\left(4G_N s\right)^{\frac{d-3}{d-2}}~.
\end{equation}
Furthermore, we notice that $\frac{\partial p}{\partial y}$ is proportional to $k$:
\begin{equation}
    \frac{\partial p}{\partial y}=\frac{ky}{8\pi G_{N} l}\left(4G_N s\right)^{\frac{d-3}{d-2}}~.
\end{equation}
Consequently, this term vanishes for a charged black brane, as expected, since flat spacetime does not introduce any additional geometrical scale. Through straightforward algebraic manipulations, one can verify that the generalized Euler equation holds:
\begin{equation}\label{localSmarrLaw5}
    \epsilon + p = sT + \mu q + y \frac{\partial p}{\partial y}~.
\end{equation}
Furthermore, substituting the expressions for $y$ and $\mu$ from (\ref{s,u,y}), one can confirm that $q$\footnote{Let us emphasize that $\tilde{q}$ denotes the constant appearing in the black hole line element, whereas $q = q(x)$ is a local function defined on the boundary spacetime. We apologize for any notational inconvenience this may cause.} and $\partial_y p$ satisfy their respective conservation laws, given by (\ref{eq:U(1)conservation}) and (\ref{eq:NEWconservation}).

To this point, we have utilized an arbitrary conformal factor to represent a curved physical background, simulating a state near out of equilibrium for the fluid. The usual global thermal state is obtained by selecting $W = Rl$, where $R$ is a constant factor, reducing equation (\ref{localSmarrLaw5}) to the Smarr formula given in (\ref{SmarrK}). Strictly speaking, the last statement is true for $R=1$; however, the extra scale allows us to distinguish the radius $Rl$ of the transverse space $\Sigma_k$ from the AdS radius $l$. 

In conclusion, we have demonstrated that the Smarr formula is a specific instance of the generalized Euler equation (\ref{localSmarrLaw5}). This result implies that holographic fluids satisfying the Smarr formula are not fundamentally distinct from more conventional fluids. The key insight is that when an additional scale arises from the underlying geometry of the thermal system, the standard scaling argument used to derive the Euler equation no longer holds, necessitating a generalized formulation. For a system with a compact volume, the physical interpretation is quite clear: the correction term $y \frac{\partial p}{\partial y}$ in the Euler equation naturally accounts for a finite-volume effect, as previously noted by \cite{Karch}. In summary, there is nothing intrinsically special about the Smarr formula for AdS black holes; it emerges naturally from general thermodynamic and fluid considerations.

%%%%%%%%%%%%%%%%%%%%%%%%%%%%%%%%%%%%%%%%%%%%%%%%%%%
%%%%%%%%%%%%%%%%%%%%%%%%%%%%%%%%%%%%%%%%%%%%%%%%%%%
\section{Comments on cosmological constant as a thermodynamic variable}\label{sec:EBHT}
%%%%%%%%%%%%%%%%%%%%%%%%%%%%%%%%%%%%%%%%%%%%%%%%%%%
%%%%%%%%%%%%%%%%%%%%%%%%%%%%%%%%%%%%%%%%%%%%%%%%%%%

In the previous sections, we argued and demonstrated that the Smarr formula is a specific case of the generalized Euler equation, particularly in the context of black holes in AdS. In this sense, the Smarr formula naturally fits within the framework of the AdS/CFT correspondence, as its origins are rooted in well-established physics. This conclusion is supported by the fact that we derived the generalized Euler equation using effective field theory techniques applied to fluids. Only after establishing this result did we invoke the AdS/CFT correspondence to interpret the Smarr formula.

Extended black hole thermodynamics was initially proposed outside the context of gauge/gravity duality to provide an interpretation of the Smarr formula. In the original version of this proposal, the cosmological constant $\Lambda$ was identified with a thermodynamic pressure \cite{Kastor, Gibbons, Dolan1, Dolan2, Dolan3} as follows:
\begin{equation}\label{CosmologicalPressure}
    p_{\Lambda}=-\frac{\Lambda}{8\pi G_{N}}=\frac{(d-1)(d-2)}{16\pi G_N l^2}~,
\end{equation}
where $d$ is the dimension of the bulk spacetime.\footnote{For reviews on extended black hole thermodynamics, see \cite{Mann4, Mann5}. Other relevant literature on the subject includes \cite{Mann2, Mann3, Clifford2, Kastor1}.} Clearly, the notion of pressure  $p_{\Lambda}$ does not correspond to the pressure of the dual field theory. For this reason, the physical interpretation of the Smarr formula was investigated from a holographic perspective in \cite{Karch}. 

Following the work of \cite{Karch}, a revised version of extended black hole thermodynamics was proposed to reinterpret the Smarr formula consistently with the AdS/CFT correspondence. To understand this new proposal, we recall the holographic identification between the parameters of gravity and those of the dual field theory:
\begin{equation}\label{HolographicDictionary}
    N^2\sim \frac{l^{d-2}}{G_{N}}~.
\end{equation}
Here, $N$ represents the number of degrees of freedom on field theory side of the duality. For instance, in $\mathcal{N}=4$ SYM, $N$ corresponds to the rank of the gauge group $SU(N)$, which is also the size of a matrix $A$ in the adjoint representation. In the revised version of extended black hole thermodynamics, the so-called central charge $C\sim N^2$ is treated as a thermodynamic variable associated with the cosmological constant $\Lambda$ \cite{Visser1, Gao, Visser2, Mann6}. In this view, a variation in the cosmological constant induces a flow in theory space, affecting the thermodynamics of black holes in AdS. This proposal is supported by a possible origin of extended black hole thermodynamics in higher-dimensional physics \cite{Pedraza}.

The main result of this paper—that the Smarr formula is a specific instance of the generalized Euler equation—conflicts this new interpretation for extended black hole thermodynamics. The reasoning is as follows: perfect fluids, a particular case of hydrodynamics, differs fundamentally from renormalization group (RG) physics. While the number of degrees of freedom $N$ can indeed change with the energy scale in RG flow (as is well understood in the context of AdS/CFT, see \cite{Ioannis1,Ioannis2}), this cannot explain the Smarr formula. In a perfect fluids, thermodynamic variables must be functions of spacetime to describe local thermal equilibrium, something that the number of degrees of freedom $N$ cannot do. The point is that we should not confuse a flow in a physical space where the fluid is on with a flow in theory space where an RG flow takes place. Finally, it is crucial to recognize that global equilibrium is merely a special case of local equilibrium. As a result, there is no justification for treating $C\sim N^2$ as a thermodynamic variable.

As a final remark for this section, the standard techniques for studying AdS black hole thermodynamics—such as the Euclidean method or your preferred method for computing conserved charges—struggle to clarify the meaning of the Smarr formula. One reason for this difficulty is the double role played by the AdS radius $l$. On one hand, $l$ can be interpreted as the radius of the transverse space at the boundary, while on the other, it sets the scale for constructing dimensionless quantities in holography, such as in equation (\ref{HolographicDictionary}). This point can be a source of confusion. However, in our construction, this issue is resolved because the variable $y$ (as defined in equation (\ref{y}) is diffeomorphism-invariant with respect to physical metric $g_{ab}$ and gauge-invariant under the fluid transformations (\ref{TemporalShift}), (\ref{spatialDiffeo}) from the outset. Thus, there is no ambiguity. Ultimately, the insights gained from fluid/gravity duality were instrumental in revealing the correct understanding of the Smarr formula within holography.

\section{Discussion}\label{Discussion}

In this work, we derived a generalized Euler equation (\ref{SmarrLaw}) using effective field theory techniques, granting our result a broader range of applicability independent of the AdS/CFT correspondence. We then applied the generalized Euler equation to a specific case: a holographic fluid derived from the boundary stress tensor associated with an AdS$_d$ Reissner-Nordström black hole, from which we recovered the Smarr formula. Based on these findings, we further argue that treating the cosmological constant as a thermodynamic variable is not physically justified. This is because any thermodynamic variable associated with the AdS radius $l$ cannot become a function of the boundary spacetime, as required for local thermodynamic equilibrium.

Our construction was restricted to the zeroth-order derivative expansion of the fluid variables, since our primary goal was to generalize the Euler equation. A natural direction for future work would be to extend this analysis to higher-derivative terms in the effective action, in order to investigate whether the introduction of the new variable modifies the transport coefficients. 

It has been demonstrated that evaluating the on-shell action of spacetimes with a future event horizon (featuring a planar horizon topology) can reproduce the standard effective action presented in (\ref{EFT1}). This was achieved by setting up a double-Dirichlet boundary problem to evaluate the on-shell action \cite{Natalia1, Liu1}. It would be intriguing to extend this construction to the more general effective action (\ref{EFTACTION2}) discussed in our article. For this purpose, it is necessary to gain a deeper understanding of the variable $y$ introduced in this work.

Another intriguing avenue for exploration is the explicit construction of the thermodynamic potential $\chi$ for rotating black holes, which also exhibit a boundary stress tensor with a perfect fluid form~\cite{Clifford0, Clifford}. In this work, we focused on even boundary dimensions $d$ to avoid the complications arising from conformal anomalies. It would be valuable to develop the most general effective action for perfect fluids that can incorporate such conformal anomalies. In lower-dimensions, a notable example is the Schwarzian action, which can be interpreted as the effective action of an anomalous perfect fluid in 1d~\cite{Jensen2}. A similar structure appears in the boundary action of a non-rotating BTZ black hole with external sources~\cite{Jensen3, Kevin,Me}.

It's well-known that the Schwarzian action captures the near-horizon dynamics of near-extremal black holes \cite{Amhed,Stanford}. In contrast, our effective action given in~(\ref{EFTACTION2}) is an off-shell formulation that captures not only the one-point functions (the stress tensor), but also the Ward identities (the Navier-Stokes equations without dissipation). In this way, following closely the work of~\cite{Me}, it should be possible to derive a decoupling limit from the action (\ref{Chi}), which encodes the dynamics of a holographic charged fluid, and connect it to the one-dimensional anomalous perfect fluid of~\cite{Jensen2}. In fact, the author of~\cite{Me} already achieved this decoupling limit for a subclass of solutions of a holographic CFT$_3$, although using a different set of variables from the fluid variables.

The Smarr formula remains valid in gravity theories with higher-order curvature corrections \cite{Tomas1}. This raises an important question: can the generalized Euler equation (\ref{SmarrLaw}) accommodate these theories? If not, it would suggest that the Smarr formula once again indicates the need for further generalizations of the Euler equation. Similar questions arise in the context of the dyonic AdS black holes with a non-trivial scalar profile discussed in \cite{Sugra1, Sugra2}, which constitute solutions to gauged supergravity.

It is worth emphasizing that the isometries of the target metric $G_{MN}$, as defined in (\ref{TargetMetric}), correspond to the emergent low-energy gauge symmetries of the hydrodynamic effective action. More broadly, asymptotic symmetries at the horizon \cite{ASG1, ASG2, ASG3, ASG4, ASG5} can be interpreted as manifestations of these emergent gauge symmetries of the boundary theory. Consequently, any generalization of horizon symmetries implies the presence of new gauge symmetries in holographic effective actions. Exploring potential connections between the present work and the novel symmetries identified in \cite{Liu5} would be a compelling direction for future research.

Similarly, the isometries of the degenerate target metric  $G_{MN}$ given in (\ref{TargetMetric}) are closely related to Carrollian diffeomorphisms. Motivated by this, it should be possible to formulate an effective action for Carrollian fluids \cite{CR1, CR2, CR3, CR4, CR5, CR6, CR7, CR8}. In this context, Carrollian diffeomorphisms would serve both as gauge symmetries of the effective action and as spacetime symmetries of the fluid’s background.\footnote{This perspective aligns naturally with the Gauge/Gravity duality \cite{Liu5}, as it likely follows from taking a flat limit of the fluid. There is also a body of work exploring the connection between Carrollian fluids and horizon symmetries \cite{CR9, CR10, CR11, CR12, CR13}.} With this action principle in place, the Smarr formula for asymptotically flat black holes should follow in a manner similar to the derivation presented in this article.

\acknowledgments

R.M. would like to thank David Berenstein, Alejandra Castro, and Ioannis Papadimitriou for various discussions on related topics that, in part, inspired this project. R.M. would also like to thank Clifford Johnson and Jorge Zanelli for helpful discussions, as well as David Berenstein, Alejandra Castro, Adolfo Holguin, Sean McBride, Andrew Svesko, and Manus Visser  for their comments on the manuscript. R.M. is grateful to the Centro de Estudios Científicos, Valdivia, and Pontificia Universidad Católica de Chile, Santiago, for their hospitality while part of this manuscript was being written. R.M.'s research was supported in part by the Department of Energy under grant DE-SC 0011702, and by the Chilean Fulbright Commission and ANID through the Beca Igualdad de Oportunidades $\#$ 56180016.

\appendix

\end{document}